\documentclass[11pt,letterpaper]{article}
\usepackage[latin1]{inputenc}
\usepackage{amsmath}
\usepackage{amsfonts}
\usepackage{amssymb}
\usepackage{graphicx}
\usepackage{booktabs}
\usepackage[left=1.00in, right=1.00in, top=1.00in, bottom=1.00in]{geometry}
\usepackage{url}
\usepackage{color}
\usepackage{setspace}
\usepackage{titlesec}
\newcommand{\overbar}[1]{\mkern 1.5mu\overline{\mkern-1.5mu#1\mkern-1.5mu}\mkern 1.5mu}

\begin{document}

\title{Decoding of the Walking States and Step Rates from Cortical Electrocorticogram Signals}
\maketitle

\noindent\author{Po T. Wang$^{\text{a}}$, Colin M. McCrimmon$^{\text{a}}$, Susan J. Shaw$^{\text{bcd}}$, Hui Gong$^{\text{bd}}$, Luis A. Chui$^{\text{e}}$, Payam Heydari$^{\text{f}}$, Charles Y. Liu$^{\text{dgh}}$, An H. Do$^{\text{e*}}$}, Zoran Nenadic$^{\text{af*}}$\\

\noindent a. Department of Biomedical Engineering, University of California, Irvine, CA 92697, USA.

\noindent b. Department of Neurology, Rancho Los Amigos National Rehabilitation Center, 7601 East Imperial Highway, Downey, CA 90242, USA.

\noindent c. Department of Neurology, University of Southern California, Los Angeles, CA 90089, USA

\noindent d. Center for NeuroRestoration, University of Southern California, Los Angeles, CA 90089, USA

\noindent e. Department of Neurology, University of California, Irvine, CA 92697, USA.

\noindent f. Department of Electrical Engineering and Computer Science, University of California, Irvine, CA 92697, USA

\noindent g. Department of Neurosurgery, Rancho Los Amigos National Rehabilitation Center, 7601 East Imperial Highway, Downey, CA 90242, USA

\noindent h. Department of Neurosurgery, University of Southern California, Los Angeles, CA 90089, USA

\noindent * Corresponding authors. E-mail: and@uci.edu, znenadic@uci.edu

\newpage

\section*{Abstract}
Brain-computer interfaces (BCIs) have shown promising results in restoring motor function to individuals with spinal cord injury. These systems have traditionally focused on the restoration of upper extremity function; however, the lower extremities have received relatively little attention. Early feasibility studies used noninvasive electroencephalogram (EEG)-based BCIs to restore walking function to people with paraplegia. However, the limited spatiotemporal resolution of EEG signals restricted the application of these BCIs to elementary gait tasks, such as the initiation and termination of walking. To restore more complex gait functions, BCIs must accurately decode additional degrees of freedom from brain signals. In this study, we used subdurally recorded electrocorticogram (ECoG) signals from able-bodied subjects to design a decoder capable of predicting the walking state and step rate information. We recorded ECoG signals from the motor cortices of two individuals as they walked on a treadmill at different speeds. Our offline analysis demonstrated that the state information could be decoded from $>$16 minutes of ECoG data with an unprecedented accuracy of 99.8\%. Additionally, using a Bayesian filter approach, we achieved an average correlation coefficient between the decoded and true step rates of 0.934. When combined, these decoders may yield decoding accuracies sufficient to safely operate present-day walking prostheses.

\section*{Keywords}
\noindent Electrocorticography; Motor cortex; Brain-computer interface; Gait; Decoding

\section{Introduction} \label{sec:intro}

Gait impairment or complete loss of gait function are common consequences of chronic spinal cord injury (SCI), with a profoundly negative impact on independence and quality of life of those affected~\cite{NSCISC2013}. In addition, many of people with SCI are wheelchair bound, which increases the risk of medical complications such as the formation of pressure ulcers or life-threatening blood clots. Surveys show that those with paraplegia due to SCI regard the restoration of walking as a high-priority rehabilitation goal~\cite{Anderson2004,Collinger2013}. Restoring able-body-like gait function to these individuals would greatly improve their quality of life, while reducing the incidence of medical complications and healthcare costs. However, there are currently no biomedical solutions capable of achieving this goal, and therefore novel approaches to this problem are needed.

Cellular approaches have shown promise in preclinical studies~\cite{hskeirstead:05}, however, these results are yet to be replicated in humans. Past and present clinical trials have focused mainly on the safety of stem cell therapy in those with acute SCI~\cite{geron,asterias}. Even if proven safe, establishing the efficacy of these approaches through definitive clinical trials may still be years away. In addition, it remains unclear whether potential findings will generalize to those with chronic SCI, which account for a large majority of SCI population. On the other hand, recent neuromodulation-based studies report on the ability to restore volitional walking to those with motor-complete chronic SCI~\cite{mlgill:18,caangeli:18}. These neuromodulatory approaches rely on epidural delivery of electrical stimulation to the spinal cord, which presumably transforms the spinal circuits into a functional state that is responsive to supraspinal commands. While promising, this approach did not succeed in all study participants and they still required a significant amount of assistance in order to achieve overground walking.

Another promising approach is to integrate brain-computer interface (BCI) systems with lower extremity prostheses, such as robotic gait exoskeletons~\cite{rewalk,ekso} or functional electrical stimulation (FES) devices~\cite{parastep}. These prostheses are normally operated by tilting the upper body or pressing buttons to initiate individual steps. When integrated with sensorimotor-rhythm-based BCIs~\cite{ceking:15}, they can be controlled in a more intuitive manner, thus emulating able-body-like function. Closed-loop operation of such an integrated system can also exploit the neuroplasticity of residual pathways between the brain and spinal motor pools~\cite{bhdobkin:93}, and in turn promote neurological recovery~\cite{arcdonati:16}.                

Preliminary studies on BCIs for the restoration of walking focused on electroencephalogram (EEG) as a source of control signals. In a series of experiments, we demonstrated the feasibility of an EEG-based BCI for walking in a virtual reality environment in both able-bodied participants~\cite{ptwang:10,ptwang:12} and those with SCI~\cite{ptwang:12,ceking:13}. Subsequently, this BCI system was used to successfully restore walking to an individual with paraplegia due to SCI, first over a treadmill~\cite{ahdo:13b} and then in overground conditions~\cite{ceking:15}. However, the relatively poor spatiotemporal resolution of EEG signals allowed only a limited number of gait parameters, such as walking/idling states, to be decoded robustly. Consequently, the application of this BCI system was restricted to elementary walk/stop tasks. To restore more complex gait functions, a BCI system must decode additional degrees of freedom from brain signals. Also, these gait parameters must be decoded with high accuracy to fulfill safety requirements.  

Prior studies have used EEG signals to decode additional gait parameters. For example,    
Presacco et al.~\cite{apresacco:11} used low-frequency EEG components for offline decoding of leg trajectories during treadmill walking. However, these low-frequency EEG bands contained the fundamental stride frequency, and so the contamination of EEG with mechanical and biological artifacts cannot be ruled out~\cite{tcastermans:14}. Furthermore, present-day commercial exoskeletons~\cite{rewalk,ekso} and FES systems~\cite{parastep} for walking do not allow trajectory control at the level of individual joints, which obviates the need for low-level trajectory decoding.  

The decoding of additional gait parameters with a sufficiently high accuracy likely requires access to invasively recorded cortical signals. A recent study by Benabid et al.~\cite{albenabid:19} reported on an individual with tetraplegia using epidural electrocorticogram (ECoG) signals to achieve BCI control of an exoskeleton for walking. However, the ECoG grids were placed over the hand motor area instead of the leg motor area, and the gait BCI operation was limited to walk/stop tasks after extensive user training.
On the other hand, our previous study has found that walking states and step rates strongly modulate subdurally recorded electrocorticogram (ECoG) signals~\cite{McCrimmon2018} naturally. Specifically, the generalized $\gamma$-band (40-200 Hz) power observed over the primary leg motor cortex (M1) is robustly increased/decreased with the initiation/termination of walking, respectively. Also, the $\gamma$ power bursts were phased-locked to individual strides across multiple walking speeds. We hypothesize that this information can be harnessed to control lower extremity prostheses, whereby the state information could be used to initiate and stop BCI walking, and the step rate information could be used to control the walking speed. To this end, the present study exploits these robust cortical signal features to design a BCI decoding algorithm suitable for the control of present-day lower extremity prostheses. Our results indicate that walking state and step rate can be decoded from ECoG signals with unprecedented accuracy.      
 
\section{Methods}\label{sec:methods}

We recruited participants with subdurally implanted ECoG electrodes for epilepsy surgery evaluation. The experiments were performed at the bedside and included placing the participants on a treadmill where they alternated between standing still (idling) and walking at three different speeds. They were instrumented with motion sensors and their leg trajectories were recorded simultaneously with their ECoG signals. These data were then used to train and validate a prediction model capable of decoding both the idling/walking state and a step rate.

\subsection{Participants}
This study was approved by the Institutional Review Boards of the University of California, Irvine and the Rancho Los Amigos National Rehabilitation Center. Participants were recruited from a population of patients temporarily implanted with subdural ECoG electrodes for epilepsy surgery evaluation. Only those who had coverage of the interhemispheric (IH) M1, i.e., the expected leg motor representation area, were included in the study. The placement of ECoG grids was determined solely based on clinical needs. 

\subsection{Signal Acquisition and Experimental Task}\label{sec:saaet}

\subsubsection{ECoG Acquisition}
We acquired ECoG data by splitting signals at the headbox of the hospital's epilepsy monitoring system. To this end, we used a 32-channel bioamplifier system (NeXus-32, Mind Media, Roermond-Herten, The Netherlands). The signals were recorded in a common average reference mode and at a sample rate of 2048 Hz. The NeXus-32 system had a built-in low-pass (anti-aliasing) filter with a corner frequency of 553 Hz. 

\subsubsection{Leg Trajectory Acquisition}
We placed a pair of L3GD20 electronic gyroscopes (STMicroelectronics, Geneva, Switzerland) on the distal femur and distal tibia to respectively measure the hip and knee trajectories. We straddled the ankle by placing a custom electrogoniometer~\cite{ptwang:11} on the dorsum of the foot to measure dorsiflexion/plantarflexion. All the sensors were attached to the leg contralateral to the ECoG grid. The leg trajectory signals were acquired by a pair of Arduino microcontroller units (Arduino Foundation, Turin, Italy).  We sent a common pulse train to NeXus-32 and the microcontroller units to synchronize ECoG and trajectory signals.

\subsubsection{Task}
\begin{figure}
	\centering
\includegraphics[width=0.4\linewidth]{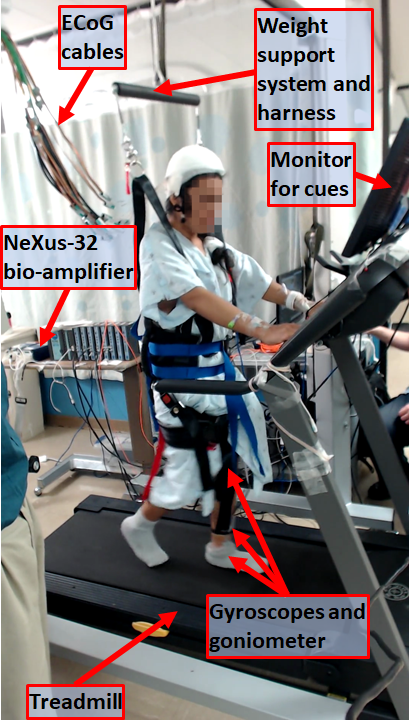}
\caption{Experimental set up showing a participant with implanted ECoG electrodes walking on a treadmill. The participant is instrumented with motion sensors to measure leg trajectories. The weight support system's only purpose is to prevent falls.}
\label{fig:setupphoto}
\end{figure}

\begin{figure}
	\centering
	\includegraphics[width=0.99\linewidth]{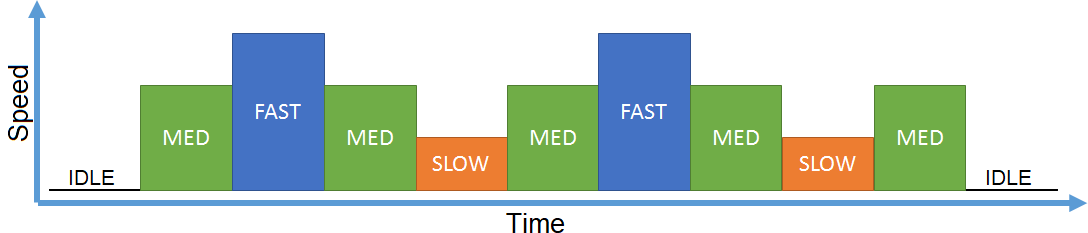}
	\caption{Design diagram for an experimental run. The nominal duration of each speed epoch is 30 s. Due to the manual adjustment of treadmill speeds, the actual epoch durations may slightly deviate from the nominal value.}
	\label{fig:expmflow}
\end{figure}

Participants walked on a treadmill (see Fig.~\ref{fig:setupphoto}) while being placed in a weight support harness (Biodex Medical Systems, Shirley, NY) to prevent falls (0\% weight support). They walked at three different speeds: Medium, Fast (50\% faster than Medium), and Slow (50\% slower than Medium), with each speed epoch nominally lasting 30 s for a total of 5 ${1/2}$ min (see Fig.~\ref{fig:expmflow}). The Medium speed was initially chosen based on gait population studies~\cite{Bohannon1997} and then adjusted so that each participant could comfortably walk at the Fast speed. The treadmill speed was adjusted manually by an experimenter. 

\subsection{BCI Decoder Overview}
\begin{figure}
	\centering
	\includegraphics[width=0.99\linewidth]{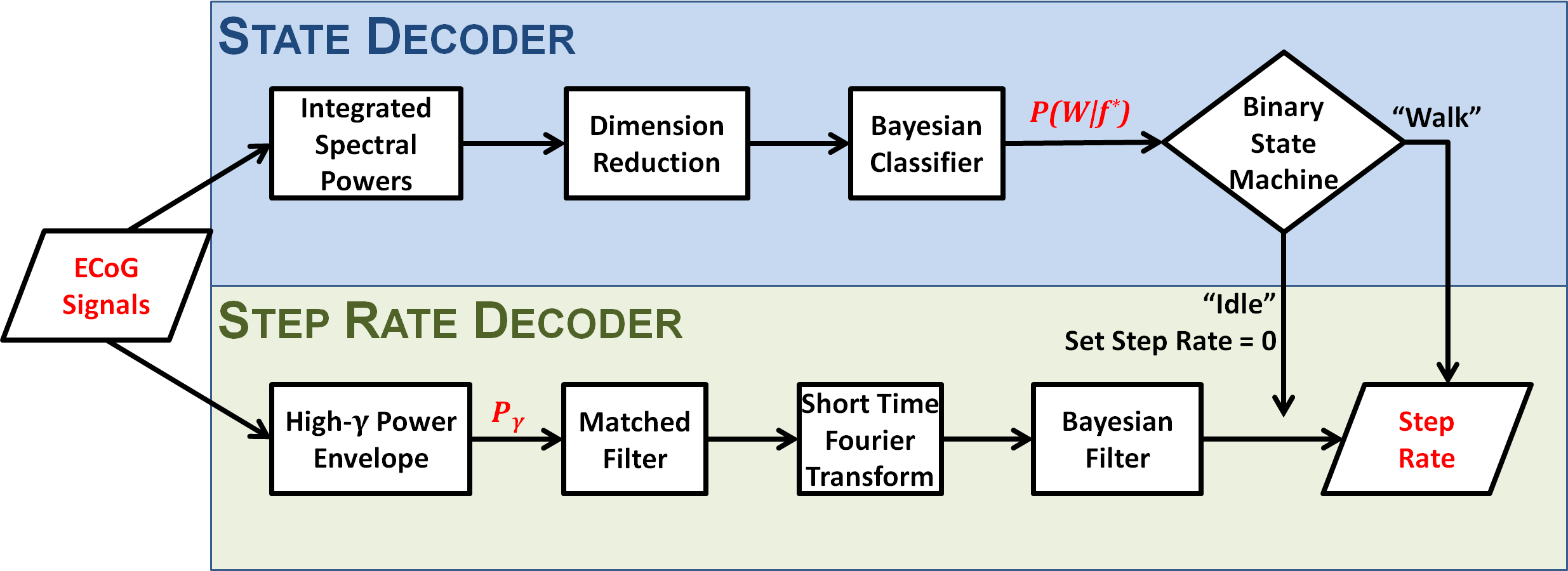}
	\caption{The schematic diagram of a two-stage BCI decoder. Each procedure will be discussed further below.}
	\label{fig:decoder_simple}
\end{figure}

We developed a two-stage BCI decoder (see Fig.~\ref{fig:decoder_simple}) as follows. In the first stage, a binary state decoder identifies whether the person is walking or not. When ECoG signals are decoded as ``not walking'' (Idle), the step rate defaults to 0. When walking is detected, the second-stage decoder applies a matched filter to the ECoG high-$\gamma$ power envelope and employs a Bayesian filter to estimate the step rate. This approach avoids the decoding of zero step rates during idling periods, which may be statistically challenging. A similar method was successfully used in our previous studies~\cite{ejputtock:13}. We trained the decoder as explained in Sections~\ref{sec:sp}~and~\ref{sec:srd}. Specifically, we used data from the first half of each experimental run and tested the decoder with the remaining half. We then repeated this procedure by reversing the roles of training and test data, and we calculated the overall performance by averaging. The decoder was implemented in MATLAB (MathWorks Inc., Natick, MA). 

\subsubsection{M1 Electrode Identification}\label{sec:mei}
We excluded electrodes outside of the IH M1 area from the analysis to ensure that the decoder only uses ECoG signals involved in gait motor control. To this end, we identified the ECoG electrode locations in the postoperative magnetic resonance imaging (MRI) scans using custom-made supervised clustering software. More specifically, we first segmented the brain from post-ECoG implantation MRI using FMRIB Software Library~\cite{Smith2002}. Next, we selected a threshold based on image intensity to isolate the ECoG electrodes in the segmented MRI brain. Since electrodes have a characteristically low intensity in MRI T1 sequence, choosing an appropriate threshold only retains voxels associated with electrodes. The segmented voxels were locally grouped by density-based spatial clustering (DBSCAN)~\cite{mester:96} and labeled by the operator. The M1 area was delineated by the central sulcus, precentral sulcus, and cingulate sulcus, and only electrodes within this region were retained for subsequent analysis. 

\subsubsection{Signal Segmentation}~\label{sec:sigseg}
An experimenter manually annotated each step based on the femur gyroscope signals to identify: 1. epochs of idling and walking; 2. epochs of the medium, slow, and fast walking; and 3. individual steps. The steps were identified using the technique described in~\cite{Aminian2002}, with the beginning of a step defined as the start of the swing phase. 
This information served as the ground truth and facilitated the BCI decoder training and testing.

\subsection{State Decoder}\label{sec:sp}
The state decoder was designed to classify whether the participant was idling or walking. We divided the leg M1 ECoG signals underlying idling and walking into contiguous, non-overlapping 750-ms segments. Subsequently, we Fourier-transformed these segments and integrated their spectral powers over the $\beta$ (20--30 Hz), low-$\gamma$ (40--55 Hz), and high-$\gamma$ (70--160 Hz) band. The $\mu$ band was excluded since it did not exhibit intra-stride modulation~\cite{McCrimmon2018} and generally contributed little to the decoding accuracy~\cite{Wang2016}. These spatio-spectral data were then converted to a log scale and subjected to dimensionality reduction by a combination of classwise principal component analysis (CPCA)~\cite{kdas:07, kdas:09} and either linear discriminant analysis (LDA) or approximate information discriminant analysis (AIDA)~\cite{kdas:08}, with final dimensions 1, 2, or 3. Finally, we calculated the posterior probabilities of Walk and Idle states for each segment using the Gaussian-based Bayes rule.

\subsubsection{Optimization}\label{sec:optimization}
We used leave-one-out cross-validation (CV) to optimize the following parameters: 1. the type of discriminant analysis; 2. the final dimension of the feature space (in case of AIDA); and 3. whether to pool the variances in the Bayes rule implementation. During the optimization process, the classifier used the maximum \textit{a posteriori} probability (MAP) rule. We ultimately chose the combination of parameters resulting in the highest CV accuracy. Note that we only used the training data in this procedure. 

\subsubsection{Binary State Machine}\label{sec:bsm}
\begin{figure}
	\centering
	\includegraphics[width=0.99\linewidth]{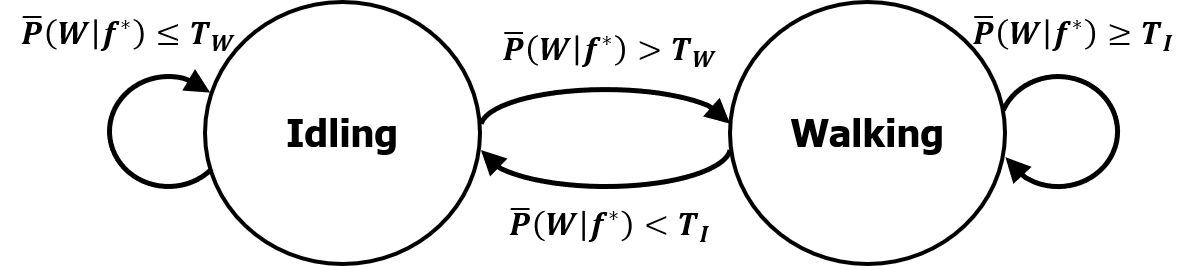}
	\caption{Binary state machine. The state transitions are marked by arrows with the corresponding transition rules next to them.}
	\label{fig:statemachine}
\end{figure}

To reduce noisy state transitions, we employed a binary state machine (BSM) with posterior probability averaging, which has been extensively tested in our real-time BCI systems~\cite{ceking:15,ahdo:13b,ahdo:11}. Specifically, at time $k$, we averaged the $n_w$ most recent posterior probabilities:
\begin{equation}\label{eq:post_average}
\overbar{P}(W|f_k^\star) = \frac{1}{n_w} \sum_{i=0}^{n_w-1} P(W|f_{k-i}^\star)
\end{equation}
where $P(W|f_{k}^\star)$ is the posterior probability of walking given features at time $k$. We used the optimal combination of parameters determined above to extract these features from a 750-ms sliding data segment (500 ms overlap) and to calculate $\overbar{P}(W|f_k^\star)$. This probability was then supplied to a BSM with the transition rules described in Fig.~\ref{fig:statemachine}. Namely, the BCI transitions from Idle to Walk state if $\overbar{P}(W|f_k^\star)>T_{W}$ and from Walk to Idle state if $\overbar{P}(W|f_k^\star)<T_{I}$, where the thresholds $T_W$ and $T_I$ satisfy: $T_W\ge T_I$.
We calibrated the BSM by seeking the highest overall in-training accuracy for all combinations of $T_{I}$ and $T_{W}$ from 25\% to 75\% in 5\% increments, as well as searching over $n_w={1,2,3}$. In the event of a tie, we selected the first tested threshold. To avoid overfitting, this procedure only used the training data.

\subsection{Step Rate Decoder}\label{sec:srd}
\begin{figure}[!htpb]
	\centering
	\includegraphics[width=0.99\linewidth]{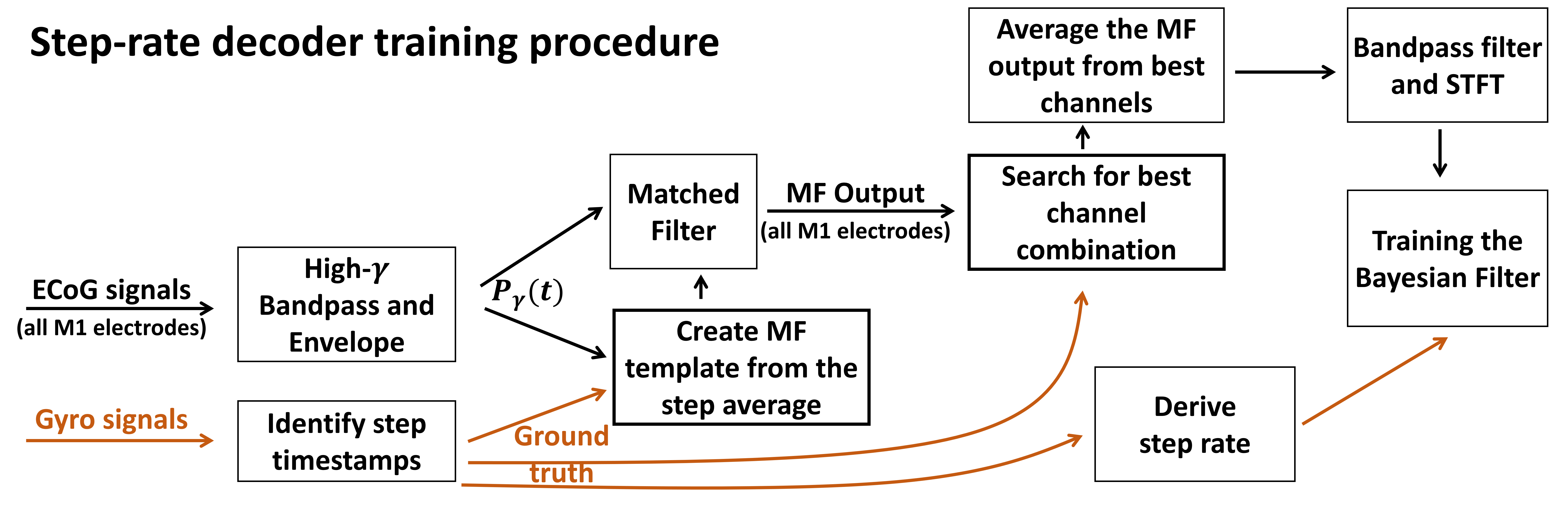}
	\caption{Flowchart for training the step rate decoder. MF = matched filter, STFT = short-time Fourier transform. Only training data were used.}
	\label{fig:trajectory_flowchart}
\end{figure}

This second-stage decoder estimated the participant's step rate whenever the BSM decoded the Walk state. The procedure for training the step rate decoder is outlined in Fig.~\ref{fig:trajectory_flowchart}. 

\subsubsection{High-$\gamma$ Power Envelope}
We designed the step rate decoder to extract stepping information from the high-$\gamma$ band. The decision to use these frequencies was informed by prior ECoG studies that demonstrated strong modulation of the $\gamma$ band with hand~\cite{crone:98,pfurtscheller:03,crone:98b}, arm~\cite{Wang2017,Ruescher2013}, as well as foot and leg~\cite{McCrimmon2018,Ruescher2013} movements. First, we band-passed (70--160 Hz) ECoG signals from all IH M1 channels using a 4th order Butterworth filter. The resulting signals were then squared, low-pass filtered (4th order Butterworth filter, corner frequency 4 Hz), downsampled to 32 Hz, and standardized using Z-score to derive the high-$\gamma$ power envelope, $P_\gamma$,~\cite{McCrimmon2018,Wang2017}. 

\subsubsection{Multi-channel Matched Filter}
We constructed a matched filter (MF) template to enhance the signal-to-noise ratio (SNR) of the $P_\gamma$ signal across all IH M1 channels. Toward this goal, we extracted $P_\gamma$ from 0.75 s before to 0.25 s after the onset of each swing phase. We empirically determined that this time window contained the most common waveform across all three walking speeds. For each channel, these 1-s-long segments were then grouped into Slow, Medium, and Fast speeds, and averaged to create a speed-specific template. The three templates were then averaged to create a multi-channel MF template. Mathematically, this resulted in a $N_t\times N_c$ matrix, where $N_t$ the number of time samples and $N_c$ is the number of IH M1 channels. 

\subsubsection{Channel Selection}
Since not all M1 channels are involved in the modulation of gait~\cite{McCrimmon2018}, including all of them could decrease the decoder performance. Therefore, we selected a subset of M1 electrodes using an optimization procedure (described below). Once the optimal subset of M1 channels was found, their MF outputs were averaged (Figure~\ref{fig:trajectory_flowchart}). 

\paragraph{Objective function} \label{sec:of}
We defined a computationally simple objective function to gauge the decoding performance when using only a subset of electrodes. This simple approach was necessary to facilitate the combinatorially complex channel selection process. Specifically, the MF outputs from a subset of channels were averaged and band-pass filtered (0.15--1.5 Hz) to remove baseline drifts and harmonics. Then, we detected peaks in this waveform and designated their arrival times as decoded step locations, i.e., the beginning of each step's swing phase. We only considered peaks above 0 (the nominal mean value of the MF output) and compared these decoded step locations to the ground truth, determined as described in Section~\ref{sec:sigseg}. If there was no decoded step location within 0.5 s of a true step location, an omission (false negative) error occurred. Similarly, a false positive error occurred if there was no true step location within 0.5 s of the decoded step location. The objective was to minimize the decoding error, $\varepsilon$, defined as the sum of omissions and false positives. 

\paragraph{Optimization procedure}
Finding the globally optimal subset of channels requires an exhaustive combinatorial search. On the other hand, choosing channels by ranking their performances could be suboptimal~\cite{tmcover:77}. Therefore, we employed the following heuristic search. First, we decoded from each IH M1 channel and calculated its corresponding decoding error $\varepsilon_i$. Channels whose decoding error was below a threshold ($\varepsilon_i<T$) participated in the subsequent combinatorial search. We defined the threshold as $T = \min\{\bar{\varepsilon},\varepsilon_{1/3}\}$, where $\bar{\varepsilon} = \frac{1}{N_c}\sum_{i=1}^{N_c} \varepsilon_i$ is the average error and $\varepsilon_{1/3}$ is the first tertile. This choice of threshold limited the number of channels in the combinatorial search. For example, outlier channels with a large error could render $\bar{\varepsilon}$ prohibitively high, in which case $\varepsilon_{1/3}$ is a more sensible threshold choice. On the other hand, a few good channels could result in $\bar{\varepsilon}<\varepsilon_{1/3}$. We then decoded from all combinations of up to N channels, where N is the number of participating channels. 
We defined the optimal subset as the combination of channels with the lowest decoding error $\varepsilon$.

\subsubsection{Bayes Filter Training} \label{sec:bt}
While useful for channel selection, the threshold-based peak detection in the above optimization procedure was sensitive to omissions and false positives and was therefore not suitable as an overall step rate decoding strategy. To decode the step rate more reliably, we adopted a Bayesian filtering approach instead. Since the power envelope, $P_\gamma$, exhibits bursting at the stepping frequency over a range of walking speeds~\cite{McCrimmon2018}, the filter utilized spectral features. Specifically,
\begin{equation}
p(s_k|f_{0:k})=\frac{p(f_k|s_k)\int p(s_k|s_{k-1}) p(s_{k-1} | f_{0:k-1}) ds_{k-1} }{ \int p(f_k|s_k) p(s_k|f_{0:k-1}) ds_k }	\qquad k=1,2,\cdots	\label{eq:bayesrule}
\end{equation}
where $s_k$ is the step rate at time $k$, $f_k$ is the spectral feature at time $k$ (defined below), $p(s_k|f_{0:k})$ is the posterior probability density function (PDF) based on features up to time $k$, $p(f_k|s_k)$ is the likelihood function, and $p(s_k|s_{k-1})$ is the step-rate transition PDF. We initialized the recursion~(\ref{eq:bayesrule}) by choosing $p(s_{0} | f_{0:0})$ as the least informative uniform PDF. Furthermore, when the state decoder reported Idle, the filter was placed in a background mode, i.e., $p(s_k|f_{0:k}) \triangleq p(s_{k-1}|f_{0:k-1})$, where $p(s_{k-1}|f_{0:k-1})$ is the last posterior of the Walk state. This mode is useful since the filter~(\ref{eq:bayesrule}) is not designed to work in Idle state, and so the calculated posteriors are meaningless. In addition, once Walk state resumes, this posterior is more informative than the uniform PDF. 
We constrained the filter by assuming $s_k \in[s_\text{min},\, s_\text{max}]$, where $s_\text{min}=0.16$ steps/s was chosen as one-half of the slowest nominal step rate during training, and $s_\text{max}=1.16$ steps/s was chosen to encompass the fastest nominal step rate during training. This choice of $s_\text{max}$ also precluded the presence of significant superharmonics in the spectral features (see below and Fig.~\ref{fig:stft_train_decoded} for example). Within this range, we discretized the step rate with a resolution of 0.005 steps/s.

\paragraph{Spectral Features}
The MF outputs of the optimal subset of channels were band-pass filtered (0.15--1.5 Hz), averaged over channels, and subjected to a 512-point short-time Fourier transform (STFT) with a continuously sliding 6-s boxcar window. 
For a data window at time $k$, we then found the maximum STFT amplitude and defined the feature $f_k$ as the frequency at which the maximum is attained. To maintain causality, the time $k$ was assigned to the leading edge of the data window.

\begin{figure}[!htpb]
	\centering
	\includegraphics[width=0.49\linewidth]{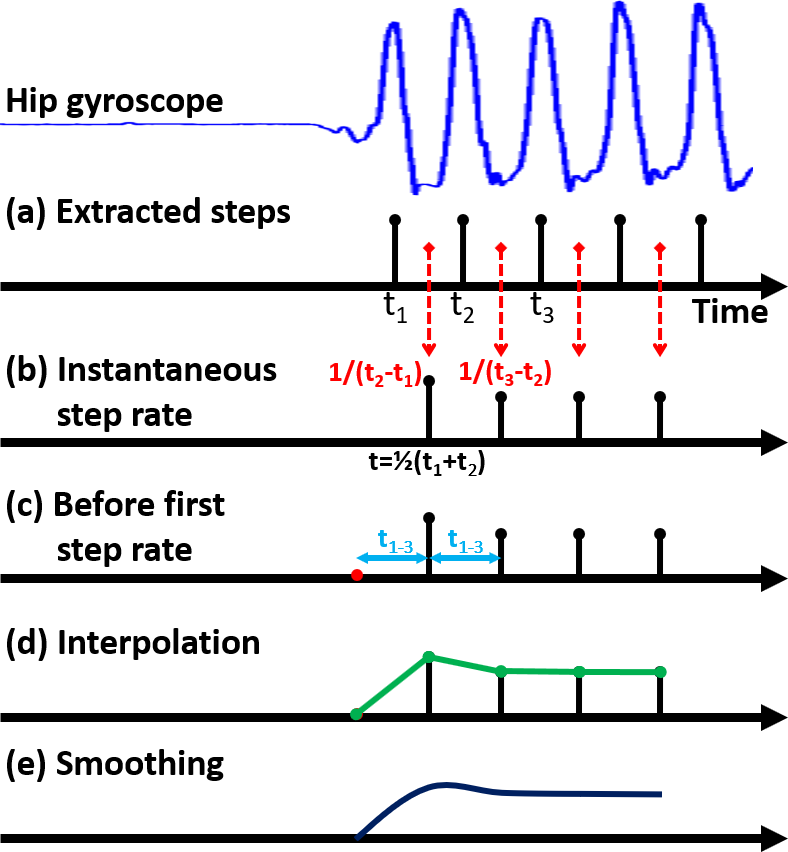}
	\caption{Ground truth step rate calculation. (a) Steps were extracted from hip and knee gyroscope signals by peak detection. (b) Each instantaneous step rate was assigned to a time point halfway between a pair of adjacent steps. (c) The ISR time series were zero-padded at the beginning (red dot) and the end (not shown). These zero step rates were added in an equidistant manner. (d) ISRs were linearly interpolated. (e) Interpolated ISRs were smoothed with a 1-s boxcar window.}
	\label{fig:steprate_truth}
	
\end{figure}

\paragraph{Ground Truth Step Rate}
We detected the peaks in the hip and knee gyroscope signals and stored their corresponding times, denoted by $t_i$. We then defined an instantaneous step rate (ISR) for the $i$th step as $(t_{i+1} - t_i)^{-1}$ and assigned it to the location halfway between $t_i$ and $t_{i+1}$. The boundary conditions and subsequent interpolation and smoothing of ISRs were handled as explained in Fig.~\ref{fig:steprate_truth}. Finally, we averaged the smoothed hip and knee ISRs to obtain a single ground truth step rate time series.

\paragraph{Likelihood Function}
By assuming that spectral features and step rates are jointly Gaussian, the likelihood function, $p(f|s)$, takes a conditional Gaussian form:
\begin{equation}\label{eqn:likelihood}
p(f | s)=\frac{1}{\sqrt{ 2 \pi \sigma_{f|s}^2 }} \exp \frac{- \left( f - \mu_{f|s} \right)^2}{2 \sigma_{f|s}^2}	
\end{equation}
where 
\begin{eqnarray*}
\mu_{f|s}		&=& \mu_f + \rho_{fs} \frac{\sigma_f}{\sigma_s} (s - \mu_s) \\
\sigma_{f|s}^2	&=& \left(1-\rho_{fs}^2\right) \sigma_f^2 
\end{eqnarray*}
and
$\mu_f$ and $\mu_s$ are the means of $f$ and $s$ respectively, $\sigma_f^2$ and $\sigma_s^2$ are the variances of $f$ and $s$ respectively, and $\rho_{fs}$ is the Pearson correlation coefficient between $f$ and $s$.

\paragraph{Step-Rate Transition PDF}
We modeled the step-rate transition PDF, $f(s_k|s_{k-1})$, as a conditional Gaussian function whose mean is found through a linear regression. Specifically,
\begin{equation}
p(s_k | s_{k-1}) = \frac{1}{ \sqrt{2 \pi \sigma_n^2} } \exp \frac{- \left( s_k - (a \cdot s_{k-1} + b) \right)^2}{2 \sigma_n^2} \qquad k=1,2,\cdots \label{eqn:transition}
\end{equation}
where $a$ and $b$ are the slope and intercept of the regression line $s_{k} = a \cdot s_{k-1} + b$, and $\sigma_n^2$ is the residual variance.

\subsection{Validation} \label{sec:v}
\begin{figure}[!htpb]
	\centering
	\includegraphics[width=0.99\linewidth]{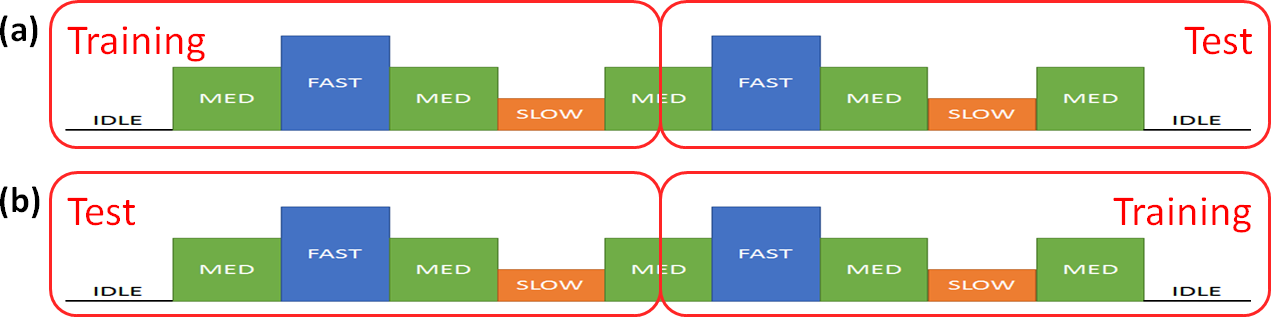}
	\caption{Validation scheme. Decoders were validated in each experimental run by splitting the data into training and test segments. (a) The first half of data was used for training and the second half for testing. (b) The roles of training and testing data were reversed.
			}
	\label{fig:validations}
\end{figure}

We used data from the first half of an experimental run to train the decoder and data from the second half to test its performance (see Fig.~\ref{fig:validations}). We then repeated this procedure by reversing the roles of the first and second halves of each run. The overall performance was obtained by averaging the decoding results over these two tests. 
In addition to default spectral features, we tested the performance of the step rate decoder by varying the STFT window length from 2 s to 5 s in 1-s increments.

We assessed the performance of the state decoder based on the number of correctly decoded Idle (Walk) segments out of the total number of ground truth Idle (Walk) segments, respectively. Additionally, we quantified the performance of the step rate decoder with the Pearson correlation coefficient, $\rho$, and root mean square error (RMSE) between the decoded and ground truth step rates during walking. Since the step rate decoder is causal, the decoded step rate is expected to lag behind the ground truth step rate. Thus, we lag-optimized the RMSE in the range $[0,\,\text{W}]$ s, where $\text{W}$ is the STFT window length.

\section{Results}\label{sec:results}
\begin{figure}[!htpb]
	\centering
	\includegraphics[width=0.99\linewidth]{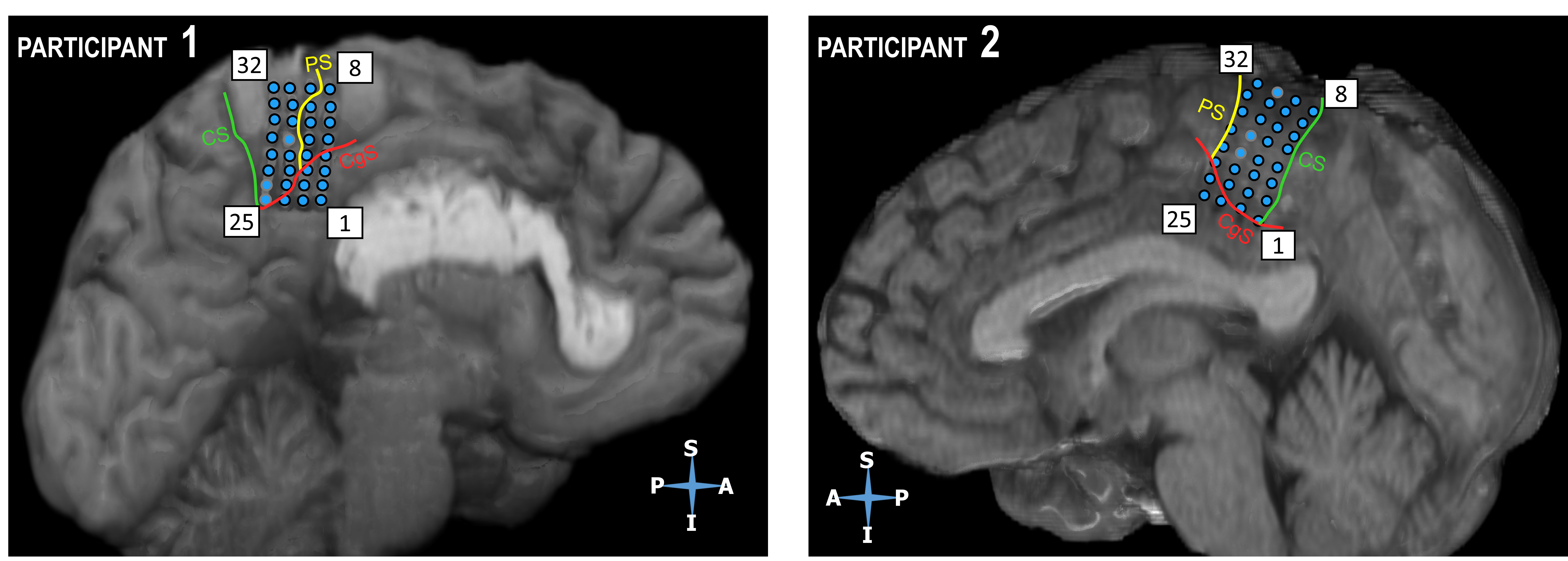}
	\caption{Placement of HD-ECoG grids. CS = Central sulcus, PS = Precentral sulcus, CgS = Cingulate sulcus. A = Anterior, P = Posterior, S = Superior, I = Inferior. Electrodes outlined in gray were disconnected due to excessive noise or technical difficulties. For Participant 2, ECoG data at electrodes G20, G21, and G24 were recovered from the hospital's monitoring system. Participants had additional grids (not shown) implanted outside the area of interest (IH M1).}
	\label{fig:ecogelectrodes}
\end{figure}

Two subjects gave their informed consent to participate in the study
Table~\ref{tab:demographics} shows their demographic data. Both participants had high-density (HD) platinum-iridium ECoG electrodes (Integra LifeSciences, Irvine, CA) implanted over the IH M1 area. Since these electrodes are smaller and have a higher density (diameter: 2 mm, pitch: 4 mm) than standard ECoG grids, they yield signals of superior spatial resolution and quality~\cite{Wang2016}. Fig.~\ref{fig:ecogelectrodes} shows the participants' grid placement. We had to disconnect a small number of electrodes due to excessive noise or other technical difficulties. For Participant 2, we replaced these missing data by those from the hospital's monitoring system. Both participants were able to complete the experimental tasks. Participant 1 had an unintended stop in the middle of the experimental run and was asked to repeat the run. Data from both runs are reported here as Run 1 and Run 2. Participant 2 had noticeable weight shifting and moved her arms during the Idle phase at the end of the run. Participant 1 had an average step rate of 0.783 steps/s (0.777 steps/s) in the first (second) run, respectively. Participant 2 had an average step rate of 0.573 steps/s. Detailed gait statistics are reported in Table~\ref{tab:gaitstats} in the Appendix. 

\begin{table}[!h]
\centering
\caption{Demographic data and ECoG grid size and location (mph = miles per hour).}
\label{tab:demographics}
\begin{tabular}{ccccc}
\toprule
Participant & Age/Sex & Grid Size & Grid Location & Treadmill Speed \\ 
\midrule
1 & 32/F  & 4$\times$8 & Left hemisphere & 1, 2, and 3 mph \\ 
2 & 38/F  & 4$\times$8 & Right hemisphere & 0.5, 1, and 1.5 mph \\ 
\bottomrule
\end{tabular}
\end{table}

\subsection{State Decoder}
\begin{figure}[!htpb]
	\centering
	\includegraphics[width=0.99\linewidth]{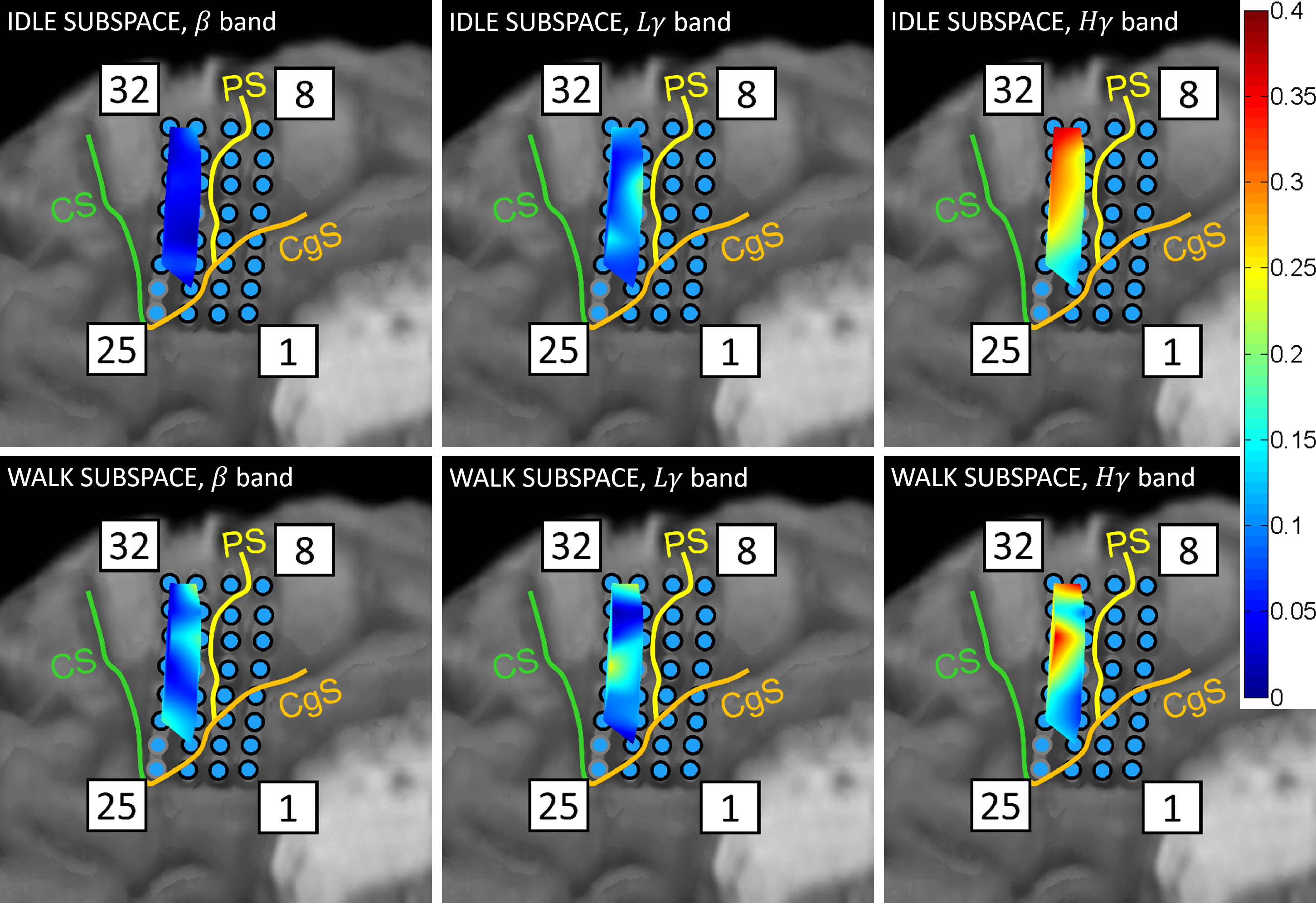}
	\caption{Representative feature extraction maps (Participant 1, Run 2) shown as the absolute values of the feature extraction matrix entries, overlaid onto the co-registered image (Fig.~\ref{fig:ecogelectrodes}). Warmer colors indicate electrode/frequency band combinations that were the most informative for the separation of Idle and Walk states. $\beta$: 20--30 Hz. $L_\gamma$: 40--55 Hz. $H_\gamma$: 70--160 Hz. Due to the piecewise linear nature of CPCA~\cite{kdas:09}, one feature extraction matrix (subspace) is generated for each state.}
	\label{fig:trfmat}
\end{figure}

Using the procedure described in Section~\ref{sec:optimization}, we found LDA and pooled variance in the Bayes rule to be optimal for both participants. Fig.~\ref{fig:trfmat} shows representative feature extraction maps, obtained by multiplying the CPCA and LDA matrices. For both participants, the most informative features were ECoG signal powers in the low-$\gamma$ and high-$\gamma$ bands, primarily in the superior part of the IH M1. The parameters of the calibrated BSM are shown in Table~\ref{tab:calibration}.

Upon calibrating the BSM, we validated the state decoder as explained in Fig.~\ref{fig:validations}. Table~\ref{tab:statevalidate} reports on the decoding results for both participants. For Participant 1, the decoding accuracy was perfect across both experimental runs. For Participant 2, the lowest performance was in the decoding of Idle state (right half). Other decoding errors included omissions in the decoding of Walk state (left half). Subsequent analysis pointed to the $\beta$ and low-$\gamma$ bands as sources of these decoding errors.
However, the exclusion of these frequency bands from the state decoder significantly lowered the overall performance. Therefore, all three bands were necessary for optimal state decoding. The average decoding accuracy  across both subjects and states was 99.8\%.  
    
\begin{table}[!h]
	\centering
	\caption{State decoder performances in the format: {number of segments correctly decoded} / {total number of segments} (percentage). The averages were weighted by the number of segments. Each segment is 750 ms with 500 ms overlap (total duration $>$16 min). Segments for which the ground truth could not be reliably determined, e.g., transitions between idling/walking and walking/idling states, or instances with gross lack of experimental compliance were excluded from this analysis.}
	\label{tab:statevalidate}
	\begin{tabular}{llr@{}c@{}@{}rrr@{}@{}c@{}rrr}
		\toprule
		Participant & Tested on & \multicolumn{4}{c}{Idle}  & \multicolumn{4}{c}{Walk} & Both \\ 
		\midrule
		1 (Run 1)	& Left half	& 158 & $\,/\,$ & 158& (100\%)	& 542 &$\,/\,$& 542& (100\%)	& 100\% \\
					& Right half	& 89 & $\,/\,$ & 89& (100\%)	& 587 &$\,/\,$& 587 & (100\%)	& 100\%  \\
		1 (Run 2)	& Left half		& 155 & $\,/\,$ & 155& (100\%)	& 541 &$\,/\,$& 541 & (100\%)	& 100\%  \\
					& Right half	& 144 & $\,/\,$ & 144& (100\%)	& 528 &$\,/\,$& 528 & (100\%)	& 100\%  \\
		2			& Left half		& 31 & $\,/\,$ & 31& (100\%)	& 564 &$\,/\,$& 568 & (99.3\%)	& 99.3\%  \\
					& Right half	& 86 & $\,/\,$ & 91& (94.5\%)	& 458 &$\,/\,$& 458 & (100\%)& 99.1\%  \\
		\midrule
		\multicolumn{5}{c}{Average} & 99.3\% 	& & & 		& 99.9\%			& 99.8\% \\
		\bottomrule
	\end{tabular}
\end{table}

\subsection{Step Rate Decoder}
\begin{figure}[!htpb]
	\centering
	\includegraphics[width=0.99\linewidth]{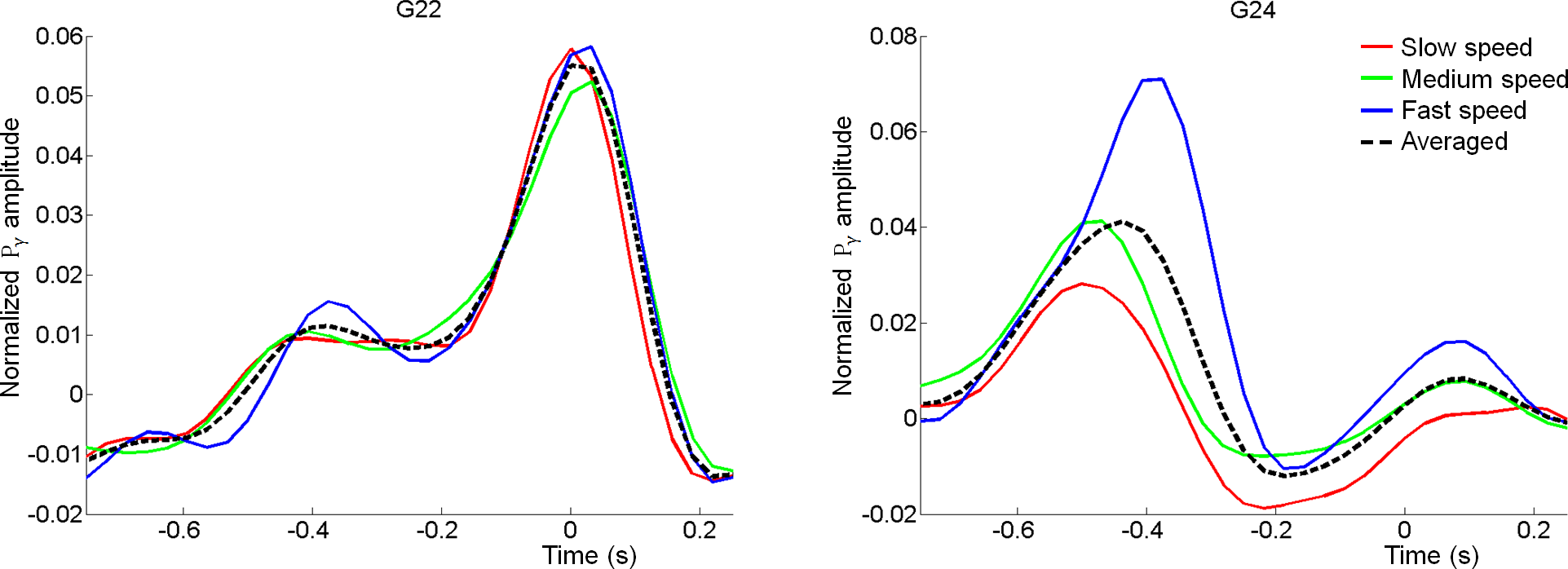}
	\caption{MF templates for two representative electrodes (Participant 1, Run 1, left half). The template averaged across speeds is shown as a dashed line. Time = 0 refers to the onset of a swing phase as defined in Section \ref{sec:sigseg}.}
	\label{fig:mftemplate}
\end{figure}

\begin{figure}[!htpb]
	\centering
	\includegraphics[width=0.99\linewidth]{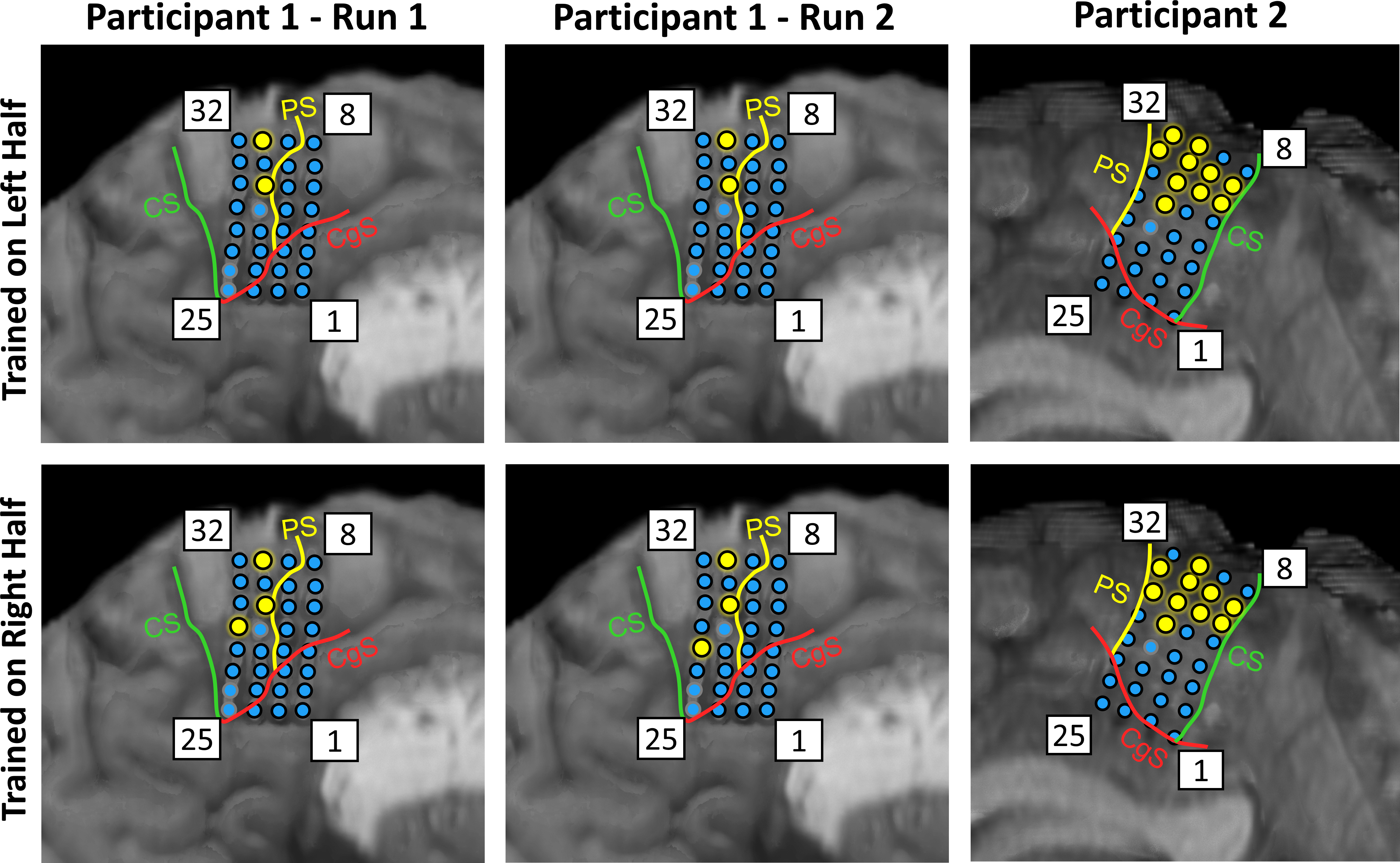}
	\caption{Optimal electrodes (shown in yellow) chosen for the decoding of the step rate.}
	\label{fig:optimalchannels}
\end{figure}

As shown in Fig.~\ref{fig:trajectory_flowchart}, the training of the step rate decoder involved the creation of MF templates, channel selection, and estimation of Bayesian filter parameters. Fig.~\ref{fig:mftemplate} shows representative examples of the MF templates. Fig.~\ref{fig:optimalchannels} shows the optimal electrodes identified through the channel selection procedure. For Participant 1, these optimal electrodes were consistent across the two experimental runs. Also, the left and right halves of each run shared two out of the three optimal electrodes.
Furthermore, the optimal electrode locations were consistent with the feature extraction maps, especially in the high-$\gamma$ band (compare to Fig.~\ref{fig:trfmat}). For Participant 2, the two halves shared nine out of 10 optimal electrodes. Similar to the feature extraction maps, the superior IH M1 was an area with the majority of optimal electrodes.    

\begin{figure}[!htpb]
	\centering
	\includegraphics[width=0.99\linewidth]{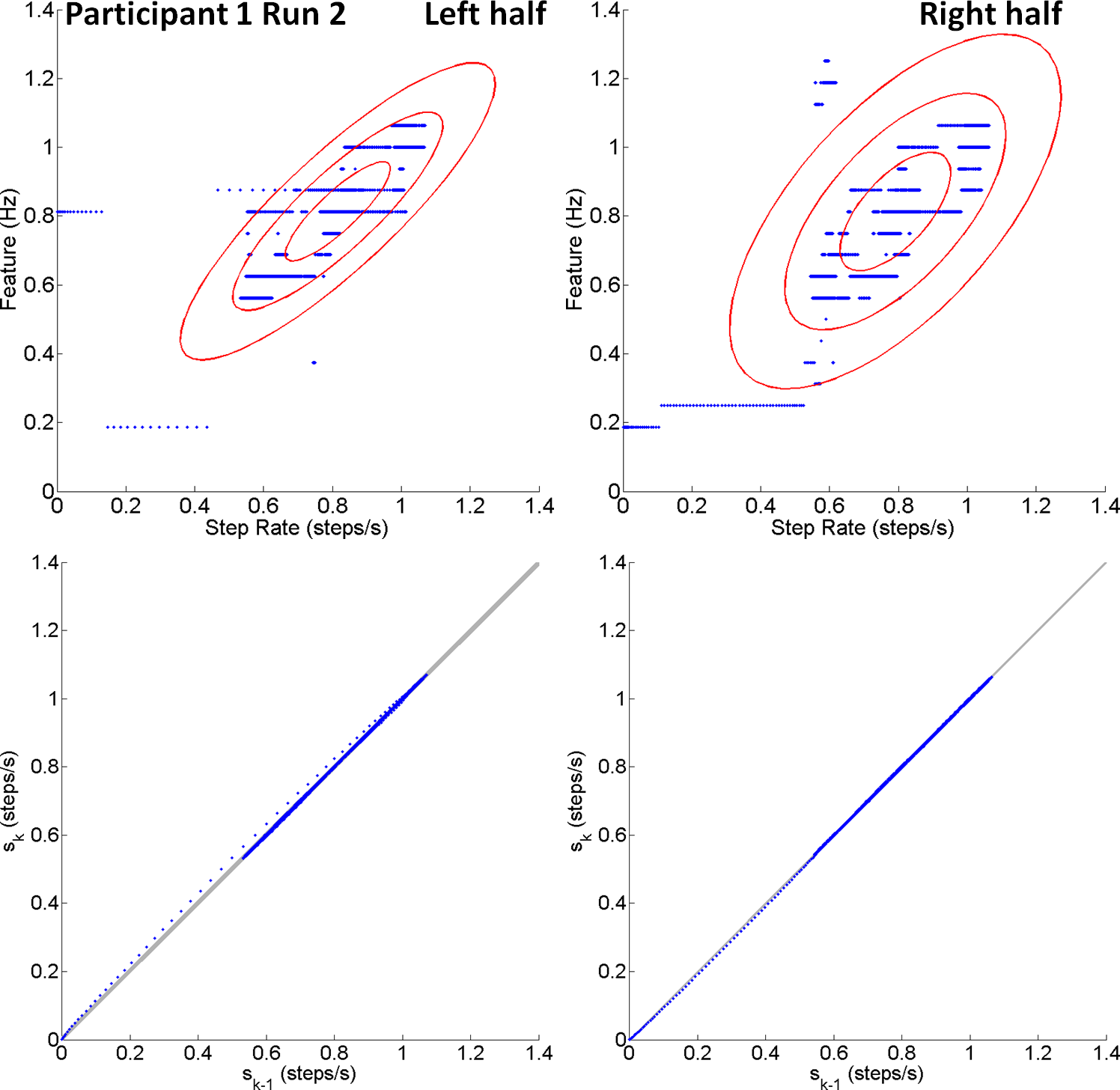}
	\caption{Representative likelihood function and transition PDF, with the corresponding training data scatter plots for Participant 1. \textbf{Top}: Likelihood function. Each sample represents a data point in the experiment with a combination of a specific ground truth step rate and feature. The feature resolution (0.0625 Hz) was imposed by using 512-point STFT on the MF output. Red concentric ellipses = \{1, 2, 3\}$\times$Mahalanobis distances.
\textbf{Bottom}: Step rate transition PDF. The gray band marks the $\pm3\sigma_n$ bounds centered around the regression line $s_k = a \,s_{k-1} + b$ (not shown). 
}
	\label{fig:liketranmap}
\end{figure}

Fig.~\ref{fig:liketranmap} shows a representative likelihood function, $p\left(f|s\right)$, and a step rate transition PDF, $p(s_{k}|s_{k-1})$, that were essential components of the Bayesian filter training, as explained in Section~\ref{sec:bt}. A prominent positive correlation in the likelihood function indicates that the step rates strongly modulate the spectral features. Additionally, since the treadmill speed remains constant during epochs of a slow, medium, and fast walking, and the transitions between these epochs are gradual, the transition PDF is highly concentrated around the regression line. The near-zero step rates in these scatter plots are due to the participant decelerating at the end of the run.

\begin{figure}[!htpb]
	\centering
	\includegraphics[width=0.99\linewidth]{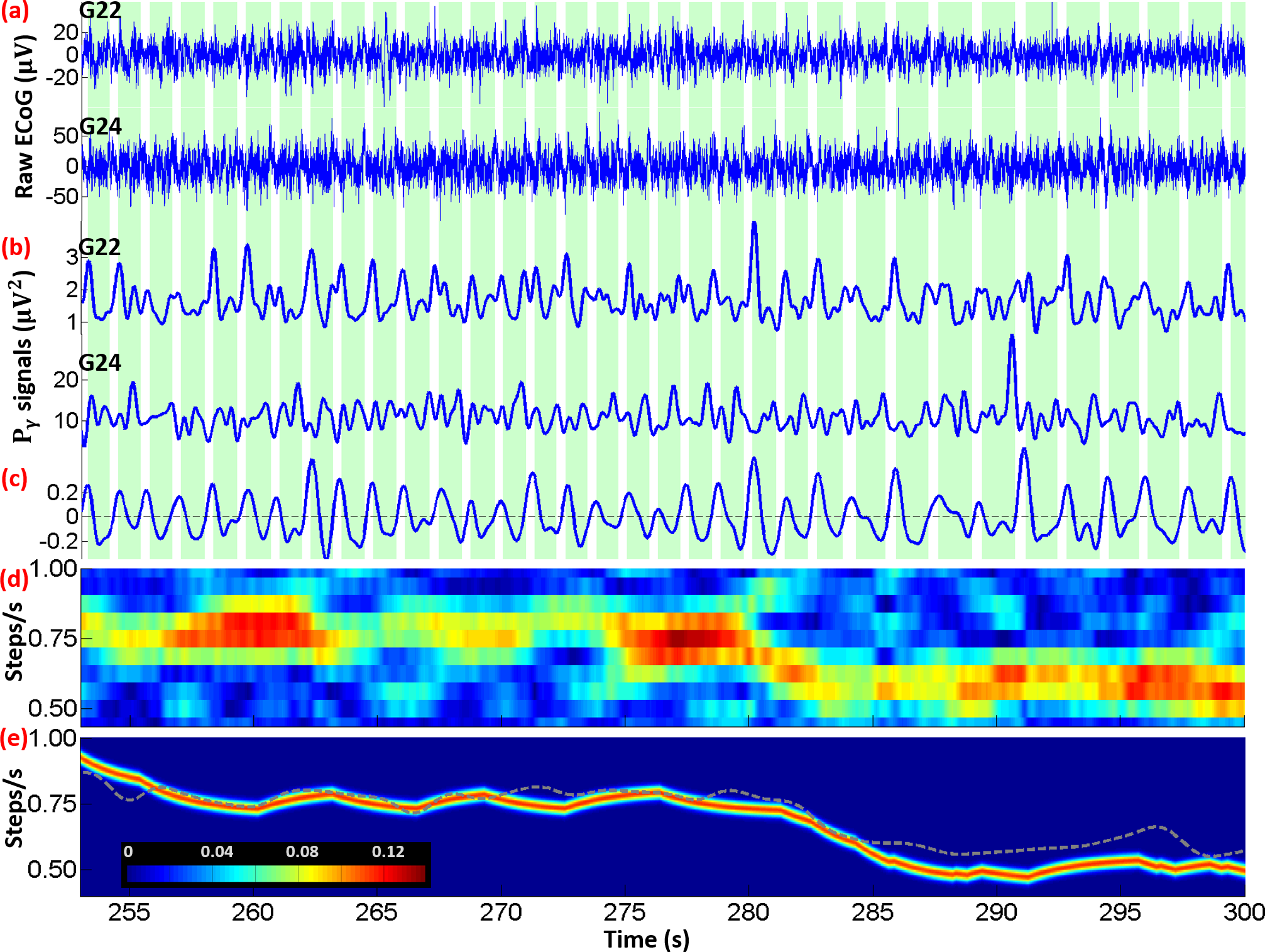}
	\caption{The outputs of the step rate decoding stages. Data are shown at the transition from Medium to Slow speed (Participant 1, Run 1, right half). The decoder was trained on the left half of the same run. (a): Raw ECoG signals at electrodes G22 and G24 (optimal channels selected by training). (b): $P_\gamma$ signals for electrodes G22 and G24. (c): MF output after averaging and band-pass filtering. Dashed line denotes zero. Green shaded boxes: Manually-identified swing phases. (d): Spectrogram of c. (e): Solid band--Posterior PDF (output of the Bayesian filter); Dashed line--Ground truth. Due to the causal nature of the decoder, the spectrogram and the posterior PDF have been shifted 4.5 s ahead for illustration purpose.}
	\label{fig:validation_multipane}
\end{figure}

We validated the step rate decoder according to the procedure described in Section~\ref{sec:v}. 
To illustrate the relationship between the physiological and kinematic data, Fig.~\ref{fig:validation_multipane} shows the representative outputs of the step rate decoding stages.
Note the clear phase-locking behavior of the MF output (panel c) where a peak was present at the beginning of each swing phase, invariant to speed.
Inspection of the MF output spectrograms indicated that the evolution of their spectral peaks over time closely resembled the ground truth step rate (see Fig.~\ref{fig:stft_train_decoded}). 

\begin{figure}[!htpb]
	\centering
	\includegraphics[width=0.99\linewidth]{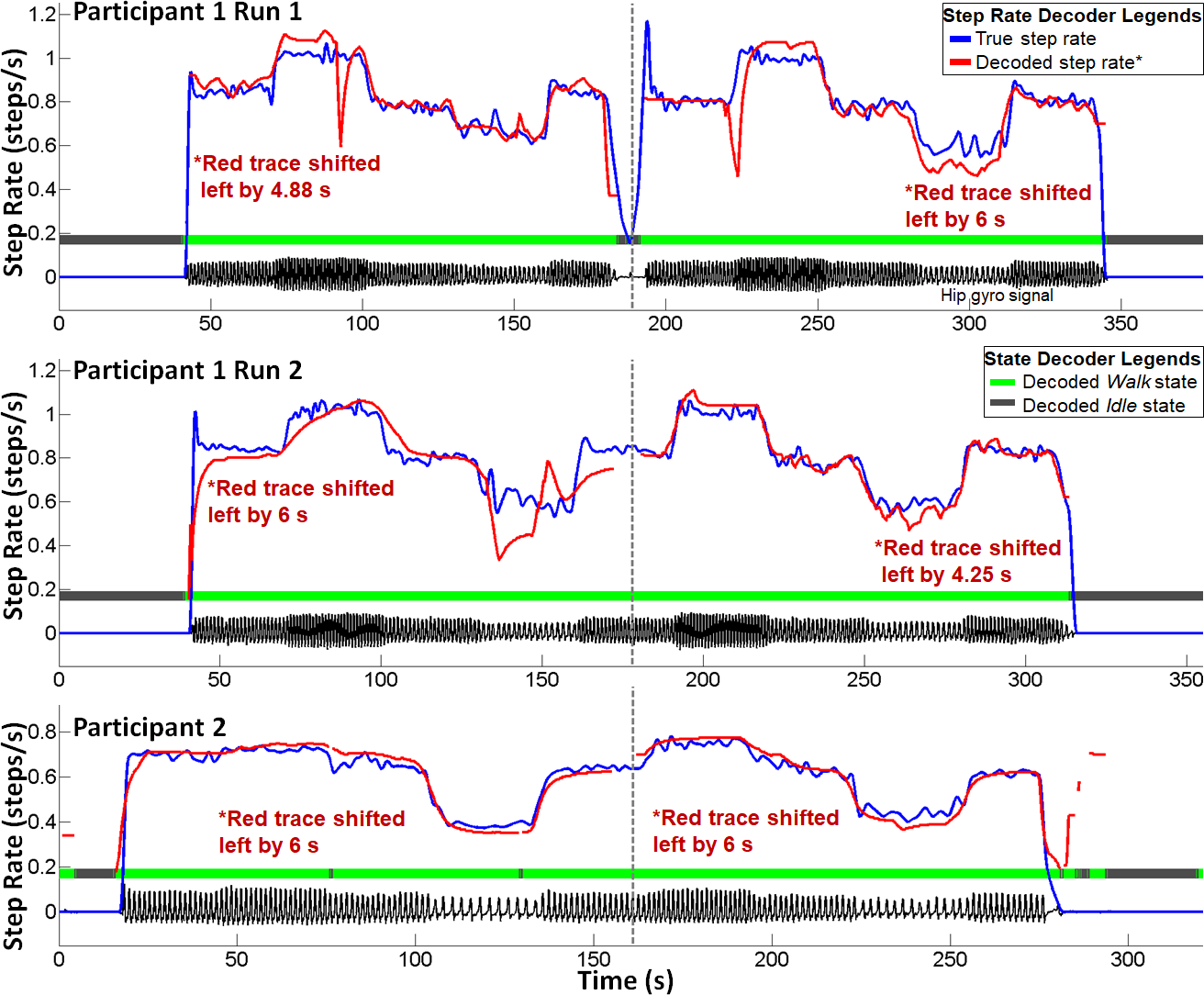}
	\caption{Decoded states and step rates. Gray dashed lines separate the left and right halves of the run. Note that 1) the red traces are shifted to the left by the optimal lag time for visual comparison; 2) when the state is decoded as Idle, the step rate decoder output is suppressed; 3) outputs are based on 6 s STFT window size.
}
	\label{fig:finalsingle}
\end{figure}

Fig.~\ref{fig:finalsingle} shows the decoded step rates obtained by combining the state and step rate decoders in the manner illustrated in Fig.~\ref{fig:decoder_simple}. As can be seen, the decoded step rates closely match the ground truth. Table~\ref{tab:steprateresult} gives a detailed breakdown of these performances. The overall average performance was $\rho$=0.934, RMSE=0.058 steps/s, and lag=5.52 s. Based on the average step rates, we translated the RMSE values into average relative errors. For Participant 1, the average relative errors in step rates were 9.3\% (Run 1) and 8.5\% (Run 2). For Participant 2, the corresponding relative error was 5.4\%. The step rate decoding performances changed when the STFT window length was varied between 2 and 6 s (see Fig.~\ref{fig:lagperf}). Interestingly, the performance improved for Participant 1 (Run 2) when the STFT window shortened to 4 s with the added benefit of a reduced lag.

\begin{table}[!h]
	\centering
	\caption{Results of the step rate decoder validation. Averages are weighted by test durations. Note that lags were optimized for lowest RMSE. } \label{tab:steprateresult}
	\begin{tabular}{llcccc}
		\toprule
		Participant	& Tested on		& Test duration & $\rho$		& RMSE		& Optimal lag \\
					&				& (s)			& 				& (steps/s) & (s) \\
		\midrule
		1 (Run 1)	& Left half		& 131.4			& 0.910			& 0.060		& 4.88\\
					& Right half	& 142.4			& 0.892			& 0.085		& 6.00\\
		\midrule
		1 (Run 2)	& Left half		& 124.9			& 0.855			& 0.095		& 6.00\\
					& Right half	& 127.8			& 0.984			& 0.038		& 4.25\\
		\midrule
		2			& Left half		& 129.5			& 0.985			& 0.029		& 6.00\\
					& Right half	& 108.7			& 0.986			& 0.033		& 6.00\\
		\midrule
		& {Average} &					& 0.934			& 0.058		& 5.52\\
		\bottomrule
	\end{tabular}
\end{table}

\section{Discussion}
To the best of our knowledge, this study represents the first attempt at decoding human gait parameters from ECoG signals subdurally recorded over the leg motor cortex.

\subsection{State decoder}
The achieved state decoding performances (99.8\%) represent a drastic improvement over similar decoders trained on EEG signals. For comparison, our prior EEG-based BCI walking studies~\cite{ptwang:12,ceking:13,ahdo:13b,ceking:15} used the identical state decoding methodology and achieved average offline performances of 77.2\%, 80.2\%, 86.3\%, and 87.0\%, respectively. Similarly, Lisi and Morimoto~\cite{glisi:15} collected EEG signals under constant and changing walking speed conditions and were able to decode these two states with an average offline accuracy of 72.7\%. The superior performance of the ECoG-based state decoder can be explained by the feature extraction maps (Fig.~\ref{fig:trfmat}), where the high-$\gamma$ band was the most informative for distinguishing between idling and walking, followed by the low-$\gamma$ band. Since these spectral features are readily accessible in ECoG signals but not in EEG, it is not surprising that EEG-based state decoders perform suboptimally. 

Our ECoG-based state decoder achieved unprecedented decoding accuracy over $>$16 min of data. Also, the decoder had a relatively small lag with a 250 ms update rate. When integrated with a lower extremity prosthesis, this type of BCI state decoder is expected to accurately control the initiation and termination of walking with an acceptable latency. 
A similar and recent study by Benabid et al.~\cite{albenabid:19} used epidurally recorded ECoG signals to decode walking states from an individual with tetraplegia. Despite having access to the $\gamma$ band, their state decoder achieved performances that were comparable or inferior to those of our EEG-based BCI systems~\cite{ceking:15,ahdo:13b}. These results could be potentially explained by a suboptimal location of ECoG electrodes, which were implanted over the arm M1 areas. Since ECoG signals in the $\gamma$ band are highly localized, their patients would be required to utilize a nonintuitive control scheme. This shows that to achieve intuitive control, e.g. the control of robotic gait prosthesis using signals from the leg M1 area, and to ``rehabilitate'' the brain areas towards a potential cellular based cure in the future, implantation at the anatomically correct location is paramount.

Long-term stability, as demonstrated by Benabid et al.'s two-year study, is also a practical concern. Even though their epidural placement could be a contributing factor in reduced decoding performance, long-term stability is a more important factor due to the cost and risk of revision surgery. Furthermore, a prior study found no significant differences between decoders utilizing signals from subdural and epidural ECoG grids~\cite{rdflint:16}.

\subsection{Step rate decoder} \label{sec:dis:srd}
The decoded step rates were highly correlated ($\rho$=0.934) with the ground truth under a variety of walking speeds. Also, the RMSE between decoded and actual step rates ranged between 5.4\% and 9.3\%. Similar to the state decoding, these performances represent a significant improvement upon the results achieved with EEG signals. For example, Presacco et al.~\cite{apresacco:11} decoded leg velocities from offline EEG data and achieved average correlation coefficients of 0.71 (ankle), 0.80 (hip), and 0.84 (knee). It should be noted that their decoder utilized low-frequency EEG components (0.1--2 Hz), thus encompassing the fundamental stepping frequency. Therefore, these performances may have been overestimated due to the contamination of EEG with mechanical and biological artifacts~\cite{tcastermans:14}. Finally, present-day commercial exoskeletons~\cite{rewalk,ekso} or FES systems~\cite{parastep} for walking are not amenable to trajectory control at the level of individual joints, which obviates the need for low-level trajectory decoding. On the other hand, they allow for speed or stepping frequency control, which is consistent with our decoding approach.

\subsubsection{Optimization}

A 6-s STFT window, used throughout this study, was sufficient to capture two steps of the instrumented leg (two strides) at the slowest walking speed (0.369 steps/s) and hence was expected to be adequate for accurately estimating the MF output spectra.
In a real-time BCI scenario, such a long analysis window would delay the system's response to user-initiated speed changes. Based on the optimal lag between the decoded and actual step rates (Fig.~\ref{fig:finalsingle}), this delay is estimated to be between 4.25 and 6 s. Shortening the STFT window could reduce this delay, albeit at the expense of lowering the correlation and increasing the RMSE (Fig. A.2).  Increased RMSE could, in turn,  lead to unsafe jerky control of a robotic gait exoskeleton. An alternative approach to shortening the response time would be to change the STFT window size adaptively in real time according to the currently decoded step rate so that the spectra of MF outputs are accurately estimated. Yet another way would be to acquire signals bilaterally, which would make both legs represented in ECoG data. This would amount to MF capturing step frequency as opposed to stride frequency, which could effectively halve the window length. Additionally, this delay is less critical for user safety, since the intention to start and stop is independent of the walking speed and can be accurately decoded every 250 ms.

\subsubsection{Sources of decoding errors}
By examining the spectrograms such as those in Fig.~\ref{fig:stft_train_decoded}, we can identify a few potential sources of error. The vertical spread around spectral peaks was $\sim$0.2 steps/s, which often led to a slight overestimation of the decoded step rate at fast walking speeds and its underestimation at slow walking speeds (Fig.~\ref{fig:finalsingle}). The spectral peak estimates could be considerably sharpened by extending the STFT window length to 8 or 10 s. However, this would result in an even greater decoder delay and, in turn, reduce responsiveness to user-initiated speed changes.

Another source of error was the weakening of spectral peaks. Fig.~\ref{fig:validation_multipane}d shows an example of this phenomenon, where the spectral peaks around 265 s decreased in amplitude, nearly reaching the noise level. This problem can be partially mitigated by the Bayesian filter since the prior probability would steer the decoder toward the most recently decoded step rate. 

In most cases, the spectral peak in a given time slice was at the frequency corresponding to the correct step rate (e.g., Fig.~\ref{fig:stft_train_decoded}). However, ECoG signals exhibited both superharmonics and subharmonics of the step rate, especially in Participant 1. These spurious spectral peaks commonly occurred at twice or one-half of the frequency corresponding to the ground truth step rate. For example, the subharmonic (0.5 steps/s) at $\sim$230 s in Fig.~\ref{fig:stft_train_decoded} caused a large decoding error. Generally, the presence of strong subharmonics resulted in a sharp temporary drop in the decoded step rate (e.g., Fig.~\ref{fig:finalsingle}), which the Bayesian filter was eventually able to correct. Subsequent analysis showed that these subharmonics were caused by a periodic alternating pattern of strong and weak $P_\gamma$ peaks. 

The presence of superharmonics could indicate that both contralateral and ipsilateral leg movements are represented in ECoG signals, which is consistent with other studies~\cite{Fujiwara2016,Ganguly2009}. We analyzed the power spectra of the $P_\gamma$ signals from the optimal electrodes and found the presence of superharmonics at twice the frequency of the step rate. This observation is consistent with our previous study~\cite{McCrimmon2018}, which found intrastride modulation of the $P_\gamma$ signals at both the fundamental stepping frequency and its superharmonics. Alternatively, these superharmonics could be associated with motor control of muscles such as rectus femoris, which is activated twice per gait cycle~\cite{tmannaswamy:99}.      

\subsubsection{Limitations and Future Work}
The main limitation of the current study is a relatively small sample size (n=2). This limitation stems from the fact that patients with intractable, medically refractory epilepsy who undergo IH ECoG grid implantation are exceedingly rare, unlike those who are implanted with grids over the cortical convexity. To test the generalizability of our findings, recruiting additional participants is needed. Since the results based on two subjects are exceedingly above the chance level, by drawing analogies to upper extremity ECoG studies~\cite{Wang2016,Wang2017}, we expect these findings to hold in a general population.   
Our future studies will recruit additional subjects to formally verify this claim, as well as investigate additional gait parameters such as changing directions and avoiding obstacles.

It is also unclear whether the current findings will ultimately generalize to individuals with SCI. However, prior studies demonstrated that decoding approaches designed based on the physiology of able-bodied individuals could be used by those with SCI for successful BCI operation. Examples include our previous studies in EEG-based BCI control for walking \cite{ahdo:13,ceking:15} and microelectrode based-BCI control of upper extremity prostheses~\cite{lrhochberg:12,taflalo:15}. Hence, there is an expectation that present findings will ultimately translate to individuals with SCI. Our future plans involve implanting volunteers with SCI with IH ECoG grids, similar to prior studies with upper~\cite{WeiWang2013} and lower~\cite{albenabid:19} extremity ECoG-based BCI prostheses.

\section{Acknowledgments}
We thank Angelica Nguyen for her assistance in setting up the experiments and Michael Chen and Aydin Kazgachi for their assistance in fabricating the gyroscopic instruments. This study was funded by the National Science Foundation 
(Award \#1446908, \#1646275). 

\section{Conflict of Interest}
The authors have no conflict of interest to disclose.

\appendix
\section*{Appendix}

\setcounter{table}{0}
\renewcommand{\thetable}{A.\arabic{table}}

\setcounter{figure}{0}
\renewcommand{\thefigure}{A.\arabic{figure}}

\begin{table}[!htbp]
	\centering
	\caption{Gait statistics derived from the sensors on the instrumented leg. Duration includes all segments of a particular target speed. Distance is calculated as treadmill speed $\times$ duration. Number of steps and average step rate are only referring to the instrumented leg. Average (Avg.) step length = Distance walked / (2 $\times$ number of steps). Average step rate = Number of steps / duration. m = meters. mph = miles per hour.} 
	\label{tab:gaitstats}
	\begin{tabular}{llrrrrr}
		\toprule
		Participant & Segment and	& Duration	& Distance	& Number of	& Avg. step	& Avg. step		\\ 
					& linear speed	& walked	& walked 	& steps		& length  	& rate 			\\
					&				& (s)		& (m)		&			& (m)		& (steps/s)		\\
		\midrule
		1 (Run 1)	& Slow (1 mph)	& 54.9		& 24.5		& 33	& 0.372		& 0.601	\\
					& Medium (2 mph)& 156.2		& 139.7		& 121	& 0.577		& 0.775	\\
					& Fast (3 mph)	& 57.0		& 76.4		& 56	& 0.683		& 0.982	\\
		\midrule
		1 (Run 2)	& Slow (1 mph)	& 60.5		& 27.1		& 35	& 0.386		& 0.579	\\
					& Medium (2 mph)& 135.3		& 121.0		& 105	& 0.576		& 0.776	\\
					& Fast (3 mph)	& 57.7		& 77.4		& 57	& 0.679		& 0.988	\\
		\midrule
		2			& Slow (0.5 mph)& 59.7		& 13.3		& 22	& 0.303		& 0.369	\\
					& Medium (1 mph)& 126.2		& 56.4		& 78	& 0.362		& 0.618	\\
					& Fast (1.5 mph)& 51.5		& 34.5		& 36	& 0.480		& 0.699	\\
		\bottomrule
	\end{tabular}
\end{table}

\begin{table}[!htbp]
	\centering
	\caption{The parameters of calibrated BSM. 
	$T_I$ and $T_W$ refer to the state transition thresholds as defined in Section~\ref{sec:bsm}. $n_w$ = number of windows for posterior probability averaging (see Eq.~\ref{eq:post_average}).}
	\label{tab:calibration}
	\begin{tabular}{lllll}
		\toprule
		Participant & Training data	& $T_I$ & $T_W$ & $n_w$ \\ 
		\midrule
		1 (Run 1)	& Left half		& 0.25 & 0.30 & 3 \\
					& Right half 	& 0.25 & 0.30 & 3 \\
		1 (Run 2)	& Left half		& 0.45 & 0.45 & 1 \\
					& Right half	& 0.25 & 0.30 & 3 \\
		2   		& Left half 	& 0.25 & 0.30 & 3 \\
					& Right half 	& 0.25 & 0.75 & 1 \\
		\bottomrule
	\end{tabular}
\end{table}

\begin{figure}[!htpb]
	\centering
	\includegraphics[width=0.99\linewidth]{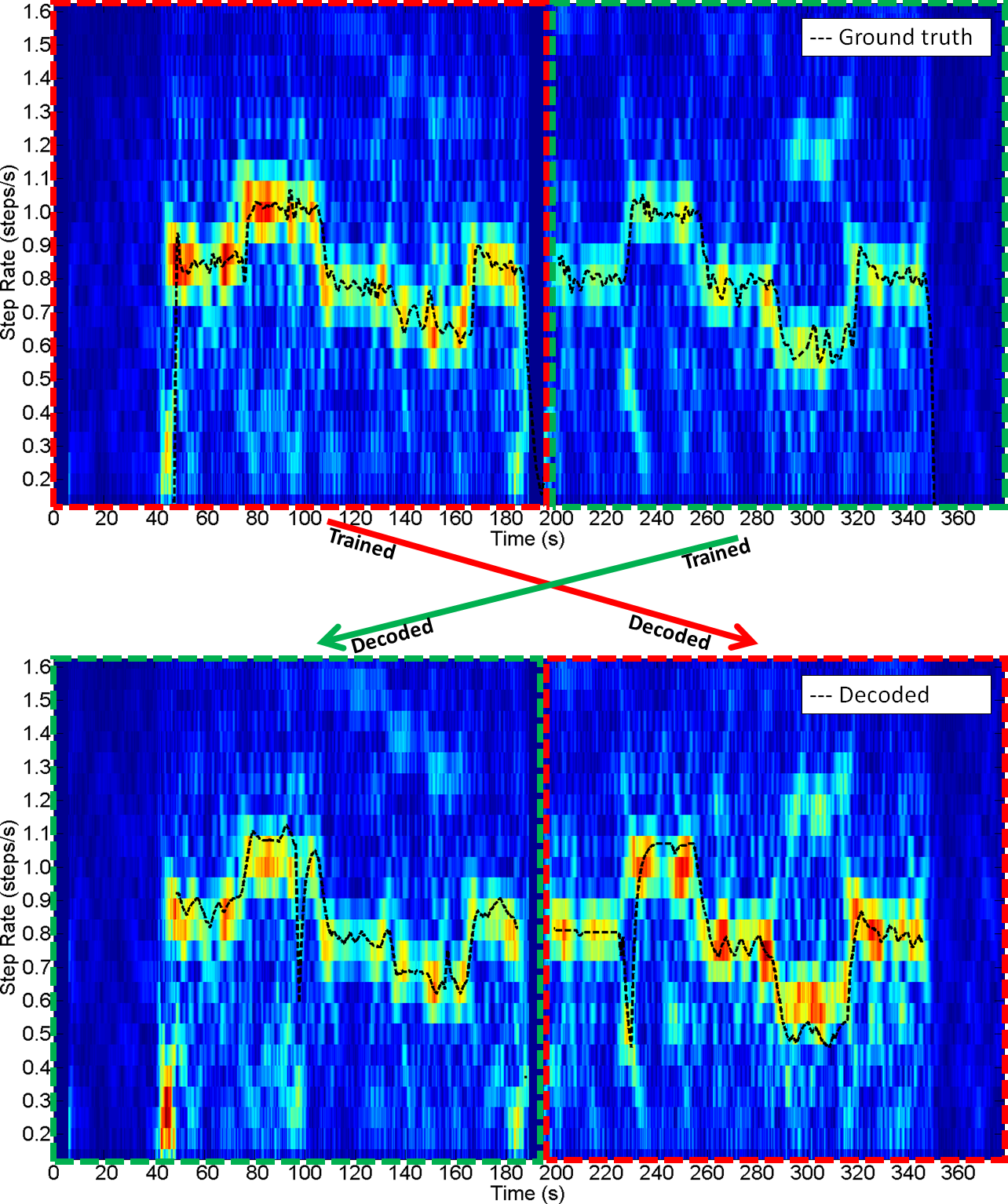}
	\caption{\textbf{Top}: Spectrogram of the MF output after averaging and band-pass filtering  for Participant 1 (Run 1). The spectrogram for each half is generated independently using the MF template constructed with data from the same half. Ground truth step rate is shifted right by 6 s to compensate for the STFT delay. \textbf{Bottom}: Spectrogram of the MF output for each data half was obtained by using template trained on the other half. Similarly, the decoded step rates for the left (right) half were obtained by the decoder trained on the right (left) half, respectively.}
	\label{fig:stft_train_decoded}
\end{figure}

The temporal evolution of the MF output spectral peaks closely resembles the step rate in both training and testing stages (Fig.~\ref{fig:stft_train_decoded}). The Bayesian filter output (bottom plot) occasionally followed the incorrect spectral peaks due to the sudden weakening of the main peak (e.g. at $\sim$95 s) or the presence of a subharmonic peak (e.g. at $\sim$230 s).

\begin{figure}[!htpb]
	\centering
	\includegraphics[width=0.99\linewidth]{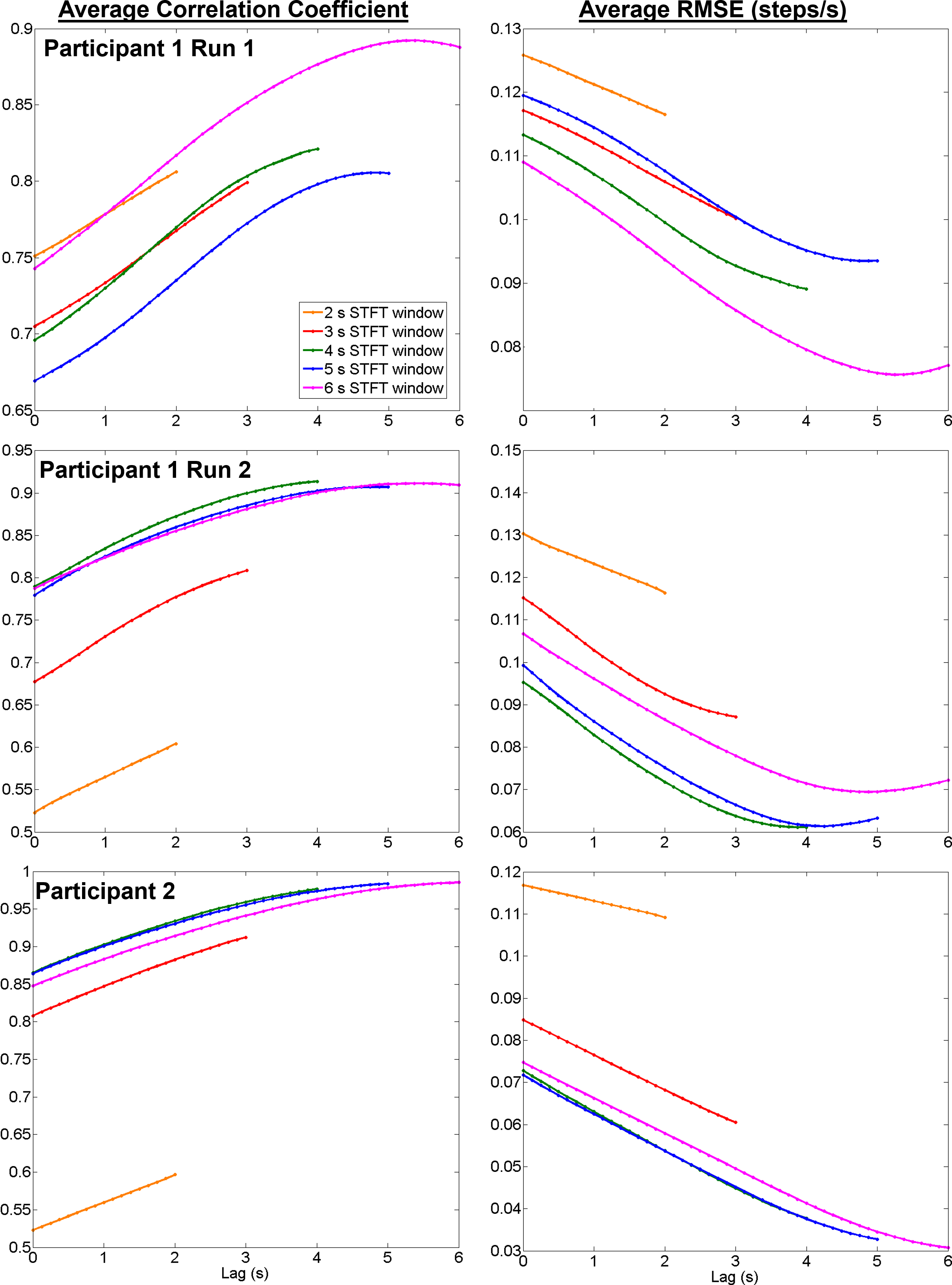}
	\caption{Decoder performances with various STFT window lengths. \textbf{Left}: Correlation coefficient vs. Lag. \textbf{Right}: RMSE vs. Lag}
	\label{fig:lagperf}
\end{figure}

\bibliographystyle{vancouver}
\bibliography{ecog_gait_decoding}

\end{document}